# Anharmonic Incommensurate Structure Modulation in Ni-Mn-Ga Martensite Exhibiting Highly Mobile Twin Boundaries

P. Veřtát[1, #], M. Klicpera[2], O. Fabelo[3], O. Heczko[1], and L. Straka[1]

[1]*FZU – Institute of Physics of the Czech Academy of Sciences, Na Slovance 1999/2, 18200 Prague 8, Czechia*
[2]*Faculty of Mathematics and Physics, Charles University, Ke Karlovu 3, 12116 Prague 2, Czechia*
[3]*Institut Laue-Langevin, 71 avenue des Martyrs, CS 20156, 38042 Grenoble cedex 9, France*

We perform neutron diffraction on a bulk $Ni_{50.0}Mn_{27.5}Ga_{22.5}$ single crystal to investigate the evolution of its five-layered modulated (10M) martensite structure, from 300 K down to 10 K. Close to martensite transformation, the modulation is nearly commensurate, with $q$ = 0.402 in a modulation vector ($q$, $q$, 0). Upon cooling, the shift in diffraction satellites indicates a transition to an incommensurate modulation with increasing $q$. However, the observed fifth-order diffraction satellites below 260 K and even more complex satellite landscape at 10 K cannot be explained by incommensurability alone. Our analysis reveals that the modulation function is highly anharmonic, encompassing Fourier components up to the eighth order. We have developed a single-parameter description of the evolution of modulation across the entire temperature interval of the 10M phase existence. Surprisingly, the structure evolution from commensurate to incommensurate modulation has no significant effect on twin boundary mobility.

# Corresponding author, *address:* FZU - Institute of Physics of the Czech Academy of Sciences, Na Slovance 1999/2, 18200 Prague 8, Czech Republic; *tel.:* +420 266 052983, *e-mail:* vertat@fzu.cz

Magnetic shape memory (MSM) alloys have attracted significant attention due to their giant, up to 12%, magnetic field-induced strain (MFIS) and their wide range of potential applications, from vibration energy harvesters to fast actuators and microfluidic pumps [1–6]. The Ni-Mn-Ga system, in particular, serves as a prototype in the MSM research field. It exhibits extraordinarily high mobility of martensite twin boundaries, especially of type II [7,8]. This high mobility is a critical precondition for MFIS occurring via twin boundary motion. Its comprehensive understanding should be based on a detailed knowledge of the crystal structure.

However, the comprehension of the crystal structure of Ni-Mn-Ga martensite remains limited and somewhat controversial. Even when focusing solely on the most studied five-layered modulated martensite, known as 10M or 5M, reports of different crystal structures emerged. Examples at the extremes are i) the monoclinic *commensurate structure* (C, $q = 2/5$) of $Ni_{50}Mn_{28.8}Ga_{22.2}$ with modulation period aligned with lattice periodicity, reported based on powder- [9,10] and high-resolution single-crystal diffraction [11], and ii) the orthorhombic *incommensurate structure* (IC, $q = 0.425$) of $Ni_{50}Mn_{25}Ga_{25}$, with modulation period nonaligned with the lattice periodicity, where the modulation vector is given in cubic ($L2_1$-derived) coordinates as $\boldsymbol{q} = (q, q, 0)$ [9,10]. Furthermore, gradual changes in the period of modulation with temperature, manifested as the changes of $q$, have been observed [10,12,13]. Consequently, the nature of the structure depends on composition, temperature, and the thermomechanical history of the sample. To date and to our knowledge, no study has specifically aimed to clarify the links between the character of modulation and the twin boundary mobility.

In this letter we investigate the evolution of the five-layered modulated structure across broad temperature range. We selected a bulk single crystal with dimensions of 0.9 mm × 2.3 mm × 10 mm and a composition of $Ni_{50}Mn_{27.5}Ga_{22.5}$ for this study. The sample exhibited the 10M structure along with exceptionally high mobility of type II twin boundaries and MFIS in the temperature range between 300 and 1.7 K. For details on these measurements, sample preparation, and basic alloy characterization, see Refs. [14,15] and Supplementary material. The aims of this study were i) obtaining an accurate understanding of the evolution of structure modulation with temperature, and ii) initiating efforts to elucidate the connections between the exact character of modulation and the high mobility of twin boundaries.

The neutron diffraction experiments were performed using the D10 four-circle neutron diffractometer at ILL Grenoble, employing a wavelength of $\lambda = 2.360$ Å [dataset][16]. Prior to the measurement, the sample was mechanically compressed to obtain a nearly single martensite variant state, i.e., to get a uniform orientation of the *c*-axis in the sample [17]. Application of bulk method of neutron diffraction on a nearly single-variant sample enabled high precision and confidence in the experimental data. After the initial sample orientation in the diffractometer, we measured the q-scans in the [110]* direction including the (2-20) and (400) reflections and their modulation satellites [18–20]. The exact character of the modulation could be then determined from the satellites locations and intensities. To account for slight changes in sample orientation during cooling or heating, we utilized an area detector. In the data processing phase, the integration area for the 1D q-scans was carefully restricted, aiming to exclude the contribution from minor martensite variants and to reduce the background noise. This is described in detail in Supplementary material.

The 10M (C) structure has a commensurate modulation with the modulation period of five atomic planes and a modulation vector *q* = ($q$, $q$, 0) with a corresponding component $q = 2/5$. The q-scans along the [110]* direction between the (2-20) and (400) reflections are shown in Fig. 1, for temperatures between 300 and 10 K. At 300 K, we observe the typical diffraction pattern of 10M (C) structure. This is characterised by a set of four equidistant satellites between the main reflections, marked as $s_1$, $s_2$, and $s_{-1}$, $s_{-2}$, located at $h = 2.4$, 2.8, 3.2, and 3.6 in Fig. 1a.

At the lowest measured temperature of 10 K, Fig. 1a, the satellite landscape appears significantly different. Notably, there is a discernible shift of the satellites indicating an increase in $q$ and a transition from commensurate to incommensurate modulation. This shift is accompanied by the emergence of additional, unexpected weak satellites, marked in Fig. 1a by red circles. Despite these changes, the four satellites are dominant and the structure still can be broadly classified as incommensurate (IC) five-layered modulated system. In order to examine the pattern evolution with greater precision, presenting the measurements on a logarithmic scale proves to be particularly advantageous, as it significantly enhances the visibility of both the prominent and weak satellites.

The q-scans upon cooling are displayed in Fig. 1b. Selected distinct cases for comparison are given in Fig. 1c. Upon cooling from 300 K, the satellites present in the original C structure begin to continuously shift and additional satellites emerge, indicating a transition to an IC structure. For instance, at 140 K (as shown in Fig. 1c), there is a noticeable increase in the number of satellites compared to the C structure at 300 K. This is evident in satellites such as $s_{-3}$, which is obscured by $s_2$ in the C structure but is distinct and adjacent to $s_2$ in the IC case at 140 K. A similar phenomenon is observed for satellites $s_3$, $s_4$, $s_5$, and $s_{-4}$, $s_{-5}$. Satellites of even higher order, such as $s_{\pm 5}$ and $s_{\pm 6}$ from (0-40) and (620) reflections, are generally indicated by green arrows in Fig. 1c.

At 10 K, the lowest measured temperature, the diffraction pattern displays further differences. The region between the (2-20) and (400) reflections appears to be divided into 24 segments, suggesting a lattice periodicity (or modulation period) of 24 atomic planes. This observation contrasts sharply with the typically assumed five or near-five atomic planes for the 10M structure. Moreover, it does not correspond to the 14M structure, which has a modulation period of seven planes, or to the NM structure, which lacks modulation. Since the observed pattern at 10 K does not align with any typical martensitic phases previously reported for Ni-Mn-Ga, our findings clearly indicate the need for a reevaluation of the existing interpretations of this material's structure.

Diffraction patterns, as illustrated separately in Fig. 1c, suggest three distinct structures. However, the pattern evolution depicted in Fig. 1b indicates that the changes are gradual rather than abrupt. We can draw two key conclusions from the neutron diffraction observations: i) From 300 K down to 10 K, the crystal structure is continuously changing, with some parameters gradually evolving as the temperature decreases; ii) the shifts in satellite positions indicate an incommensurate structure. Yet, the satellites of order higher than four are unusual since the intensity decreases significantly with increasing satellite's order [21,22]. Therefore, the extensive satellite landscape suggests complexity beyond simple incommensurate modulation.

To decipher the origins of the unique diffraction patterns with unusual additional satellites, we performed a series of calculations on a model crystal using a *DISCUS simulation package* [23]. Inspired by the complex structure of $Rb_2ZnCl_4$ [24,25], we considered two major modulation characteristics leading to four different scenarios. The modulation period could be *commensurate* or *incommensurate*, and moreover the modulation function could be either *harmonic* or *anharmonic* [21,22]. For our model case, we choose $q = 3/7 \approx 0.428$ as the limit case of the incommensurate structure [13], and the sawtooth modulation function, for its broad frequency spectrum expected to be reflected in an extensive satellite landscape. We also explored variations in the modulation function amplitude, but these details are omitted as they do not alter the conclusions obtained.

The results of the four model scenarios are summarised in Fig. 2. The first scenario, depicted in Fig. 2a, represents the *commensurate-harmonic* combination, i.e. commensurate modulation and harmonic modulation wave. In this case the diffraction pattern is very clear and essentially identical to that observed in experiment at 300 K. There are only four equidistant satellites $s_2...s_{-2}$ between the (2-20) and (400) reflections and the pattern does not show any other special features.

The second scenario is *commensurate-anharmonic,* i.e. the combination of commensurate modulation and anharmonic modulation wave, Fig. 2b. The pattern is quite similar to the case of *commensurate-harmonic*, there are again only four equidistant satellites. However, the two cases show slightly different satellite intensities.

The result for the third scenario, *incommensurate-harmonic* combination, is different from the two previous scenarios, Fig. 2c. As the modulation wave is incommensurate $q = 0.428$, the distance between the satellites becomes longer than for the commensurate case $q = 2/5$ (= 0.400). The number of satellites increases as the previously obscured satellites appear, as discussed above for the experiment and satellites $s_2$ and $s_{-3}$. However, in contrast to experiment, Fig. 1, it is apparent that the intensity of satellites decreases rapidly with satellite's order. Satellites of fourth order, $s_{\pm 4}$ are very weak, and the intensity of fifth $s_{\pm 5}$, and higher order satellites is virtually zero. In contrast, our experiment shows larger intensity of the fifth order satellites $s_{\pm 5}$ than that of $s_{\pm 3}$ and $s_{\pm 4}$. Given that the incommensurate-harmonic combination fails to replicate the experimental observations, it is inadequate to attribute all observed features exclusively to an incommensurate modulation model.

The fourth scenario, featuring the *incommensurate-anharmonic* combination, emerges as the most satisfying, Fig. 2d. Whilst satellites corresponding to $s_1...s_3$ and $s_{-1}...s_{-3}$ could be identified in the pattern, there are also many other additional satellites of lower intensity. The satellites evenly divide the space between the main reflections and are apparently result of the specific combination of incommensurate modulation and anharmonic modulation function.

The close resemblance between the calculated satellite-rich pattern in Fig. 2d and the experimental pattern in Fig. 1, especially at low temperatures, indicates that the modulation function is not only incommensurate but also anharmonic. The high-order satellites in the diffraction pattern are apparently linked with the shape (anharmonicity) of the modulation function. The challenge of explaining the unusual experimental diffraction pattern transforms to the task of determining the anharmonic modulation function shape from the intensities of the satellites.

To determine the anharmonic modulation function, we fitted calculated diffraction patterns to experimental data, extending the analysis to include harmonics up to the 12$^{th}$ order of Fourier series. This approach significantly surpasses the traditional analysis limited to 3$^{rd}$ order harmonics in prior studies on Ni-Mn-Ga. The ability to conduct such detailed analysis stemmed from observing a rich satellite landscape, a direct result of our experiment designed to achieve high intensity and resolution in the *q*-direction within reciprocal space. This design contrasts markedly with conventional single crystal or powder diffraction techniques with lower resolution and/or massive peak overlapping in case of our material. On the other hand, our measurements covered only a small segment of reciprocal space, rendering the dataset suitable for determining the modulation function but not sufficient for full crystallographic analysis.

Building on the findings of numerous studies indicating (110)[1-10]$_{L2_1}$ shifting system [7], the modulation displacement was set to be transverse to the direction of modulation. We did not consider occuppational modulation nor magnetic contribution [11], as these are absent or weak in our alloy. The nearly tetragonal lattice with $a \approx b$ and $\gamma \approx 90°$ enables a convenient description of the modulation in the direct space. The coordinate system was thus defined with $x$ along the modulation, i.e., along [110], and $y$ along the [1-10] shifting (displacement) of planes, see insets in Fig 4 for illustration. The [1-10] displacements (atom shifts) within the same (110) plane were assumed to be uniform across atom types. In reality, the displacements can differ slightly for different atoms [26], although the effect is presumed to be minor and not to significantly influence further results. The initial modulation function form was chosen as:

$$dy = A_1 \cdot \sin(2\pi x/q') + \sum_{N=2}^{N=12} k_N \cdot A_1 \cdot \sin(N \cdot 2\pi x/q' + \varphi_N) \qquad (1),$$

where $A_1$ and $k_N A_1$ represent the amplitude of the fundamental and the amplitude of the $N^{th}$ harmonics, respectively; $x$ stands for the coordinate along the modulation direction (plane number $x = 0, 1, 2, ...$), $dy$ is the plane displacement along [1-10] direction, $N = 2..12$ expresses the order of harmonic, $q'=2/q$ stands for the modulation period in atomic planes, and $\varphi_N$ is the phase of the harmonics.

By fitting, we obtained modulation function universal for all measured temperatures:

$$dy = A_1 \cdot \sin(2\pi x/q') + 0.2 \cdot A_1 \cdot \sin(5 \cdot 2\pi x/q') + 0.04 \cdot A_1 \cdot \sin(8 \cdot 2\pi x/q') \qquad (2),$$

From the fits it follows that the modulation period decreases from $q' = 4.98(1)$ planes at 300 K ($q = 2/4.98 = 0.402(1)$) towards $q' = 4.80(1)$ planes at 10 K ($q = 2/4.80 = 0.417(1)$). Simultaneously the modulation amplitude increases as the temperature decreases: $A_1$(300 K) = 27 (±1.3) pm ≈ 6% of plane spacing $d_{110}$ and $A_1$ (10 K) = 42 (±2.5) pm ≈ 10% of $d_{110}$. Two harmonics were identified: the dominant 5$^{th}$ harmonic ($k_5 = 0.2$) and the weaker 8$^{th}$ harmonic ($k_8 = 0.04$), with all phase shifts $\varphi_N$ being zero. By introducing $q = f(T)$ and $A_1 = f(q)$, we successfully fitted all diffraction patterns across the entire studied temperature range of 10–300 K using only one fit parameter, $q$. Further details on the modulation function and fits are available in the supplementary material.

The comparison of experimental and calculated diffraction patterns, employing the modulation function according to Eq. 2, is illustrated in Fig. 3. Full temperature evolution with 20 K step is given in supplementary material. The character of the experimental patterns is accurately replicated, with all notable observed satellites present in the calculated pattern. Minor differences in satellite intensities between the experimental and calculated data are observed, attributable to experimental errors and instrumental factors as well as potential imprecision in our modulation function. Nevertheless, these differences mainly occur in the weak-intensity regions, i.e. two orders weaker than the main reflections.

The structure at 300 K is nearly commensurate, displaying four satellites between main reflections as shown in Fig. 3a. The calculation, accounting for a minor deviation from commensurality, similarly predicts four satellites. At 280 K, the deviation from commensurality increases, with a modulation period of 4.92 planes ($q = 0.407$), Fig. 3b. Despite this, the experimental pattern closely resembles that at 300 K, featuring two main reflections with four intervening satellites. In contrast, the calculation indicates the presence of several additional satellites.

At 220 K, the additional satellites further separate from the main peaks and the four strongest satellites, becoming distinctly visible in the experimental pattern, Fig. 3c. At 10 K, the additional satellites are most pronounced, Fig. 3d. To our knowledge, such a complex satellite pattern has not been previously reported for Ni-Mn-Ga, although it has been observed in Ni-Mn-Ga-Fe [13].

The modulation function and plane displacements for nearly commensurate and incommensurate modulation, as determined by Eq. 2, are drawn in Figs. 4a and b, respectively. The figures also detail the individual Fourier terms of the anharmonic function and displacements in proportion to the lattice in insets. The amplitude of the $5^{th}$ harmonic is significant, accounting for 20% of the fundamental, while the amplitude of the $8^{th}$ harmonic is smaller, thus contributing less to the overall shape of the modulation function.

The (nearly) commensurate structure is not affected by the $5^{th}$ harmonic, despite its significant amplitude. The crystal planes are exactly aligned with the $5^{th}$ harmonic nodes, Fig. 4a, leading to zero displacements from this harmonic. Therefore, the displacements closely resemble the fundamental harmonic signal, rendering the detection of the anharmonicity in the commensurate structure virtually impossible [27]. When the structure changes to an incommensurate phase, the $5^{th}$ harmonic begins to play a significant role. For illustration, refer to atomic plane number 6 and blue arrows in Fig. 4a, b.

The temperature-dependent evolution of the $q$ parameter is presented in Fig. 4c, derived from fitting the experimental data shown partly in Figs 1 and 3. The $q(T)$ curve saturates at low temperatures and displays a thermal hysteresis that mirrors findings from our prior research on the Ni-Mn-Ga-Fe alloy [13]. Additionally, Fig. 4c juxtaposes this data with the temperature dependence of twin boundary mobility (twinning stress) reported in Ref. [15], indicating that twin boundaries maintain high mobility across the *q* range of 0.402(1) to 0.416(1).

In summary, our theoretical model replicates the unique and thermally evolving neutron diffraction patterns observed in $Ni_{50}Mn_{27.5}Ga_{22.5}$ single crystal. Precision of our experiment enabled us to detect also very weak satellites neglected in the majority of previous studies. The diffraction patterns depicted in Fig. 1c would normally imply the presence of three distinct crystal structures. Nevertheless, the theoretical calculations and experimental data presented in Figs. 2 and 3 revealed that these seemingly different patterns are variations of the same *anharmonic modulation function*, Eq. 2 and Fig. 4, distinguished only by minor adjustments in the modulation vector component *q* and modulation amplitude $A_1$. Using only a single parameter *q* and $A_1 = f(q)$, we successfully modeled the evolution from a simple diffraction pattern at 300 K to a complex one with numerous satellites at 10 K. From the comparison of evolution of *q* and twinning stress with temperature, Fig. 4c, we conclude that variations in *q* within 0.402 to 0.416 have minimal or no significant impact on twin boundary mobility. Question for future remains whether the determined modulation function is universal for Ni-Mn-Ga(-Fe) system with five-layer modulation and what is the physical mechanism behind the commensurate-incommensurate transition.

The authors acknowledge the Institut Laue-Langevin (ILL) for the beam time allocated [dataset] [16], the funding support from the Czech Science Foundation [grant number 21–06613S], and the assistance provided by the Ferroic Multifunctionalities project, supported by the Ministry of Education, Youth, and Sports of the Czech Republic [Project No. CZ.02.01.01/00/22_008/0004591], co-funded by the European Union.

**Fig. 1.** a) q-scans along the [110]* direction between the (2-20) and (400) reflections for temperatures of 300 K and 10 K. Indicated are satellites showing five-layer modulation ($s_1$, $s_2$, ..., $s_{-1}$, $s_{-2}$, ...) and additional satellites at 10 K (within red circles). b) q-scans showing temperature-induced variations, with intensity visualized on a logarithmic scale. c) Comparative q-scans at 300, 140, and 10 K on a logarithmic scale, illustrating the evolving modulation satellites. For measurement at 140 K, the satellites unrelated to the (2-20) or (400) reflections by the distance of $q \approx 0.4$ are marked with green arrows. These could be interpreted as $s_{\pm 5}$ and $s_{\pm 6}$ from (0-40) and (620).

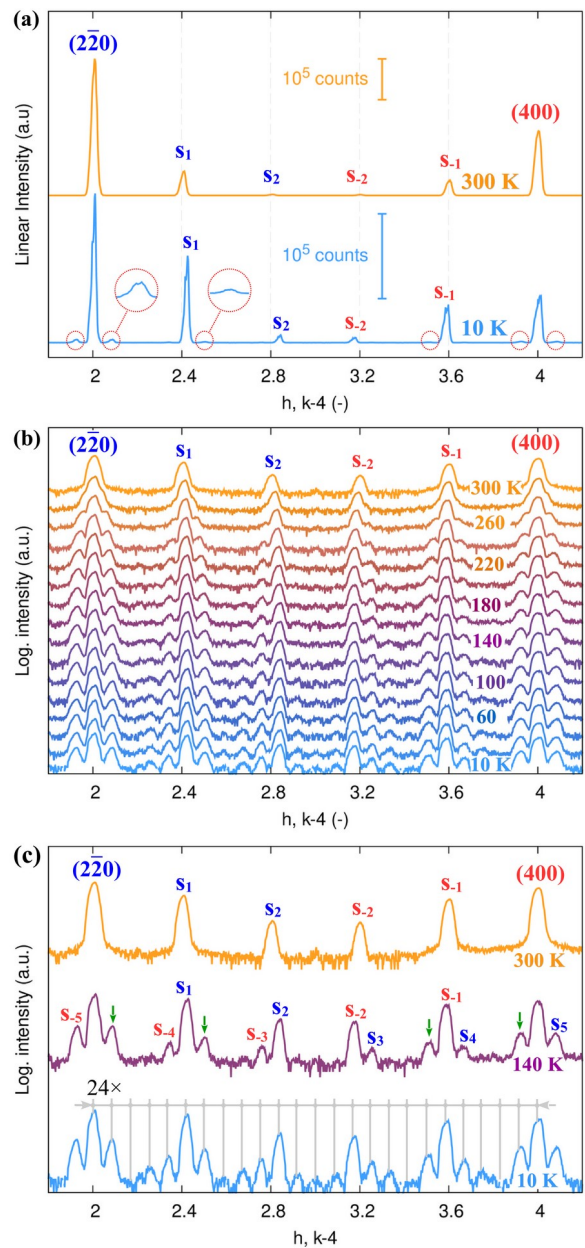

**Fig. 2.** Calculated diffraction patterns for the combinations of *harmonic/anharmonic* modulation functions and *commensurate/incommensurate* modulation periods. Main reflections and associated satellites are annotated to highlight the differences. Combinations: a) *harmonic-commensurate*, b) *anharmonic-commensurate*, c) *harmonic-incommensurate*, d) *anharmonic-incommensurate*.

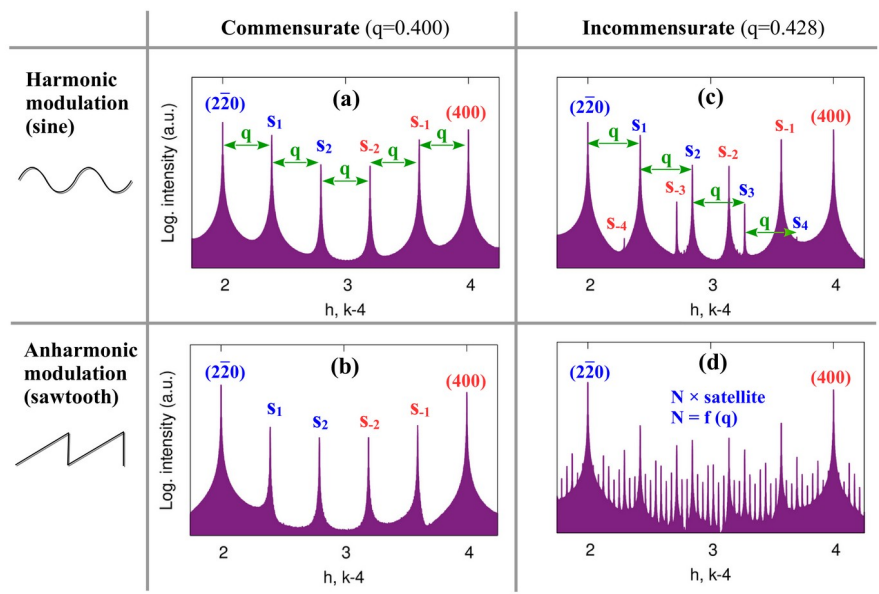

**Fig. 3.** Comparison between the observed (upon cooling) and calculated diffraction patterns for the structure model using anharmonic incommensurate modulation, demonstrating the fit between the model and experimental data. Only selected temperatures, illustrative enough to demonstrate the temperature development, are shown: a) 300 K, b) 280 K, c) 220 K, d) 10 K. See supplementary material for all measured temperatures.

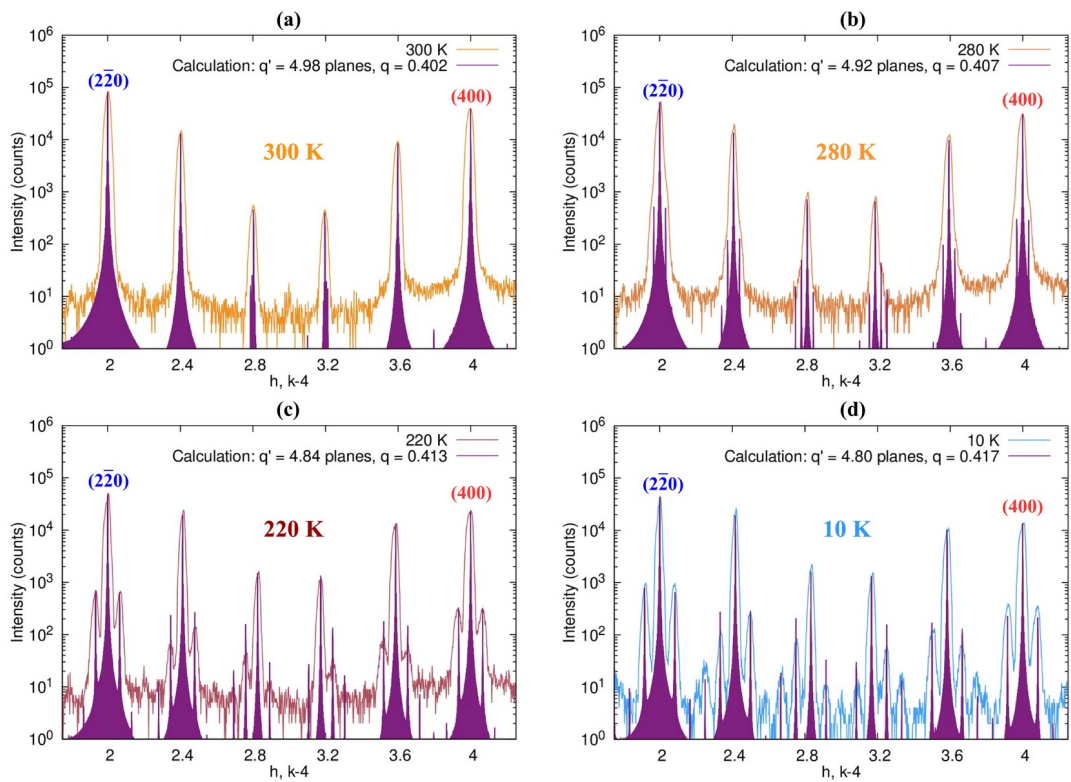

**Fig. 4.** a, b) The determined anharmonic modulation function, its Fourier components, and plane displacements for commensurate and incommensurate modulation. The insets show plane displacements in proportion to the lattice. c) Determined modulation component *q* as a model parameter together with twinning stress (from Ref. [15]) as functions of temperature. Error bars for *q* at three selected temperature are marked.

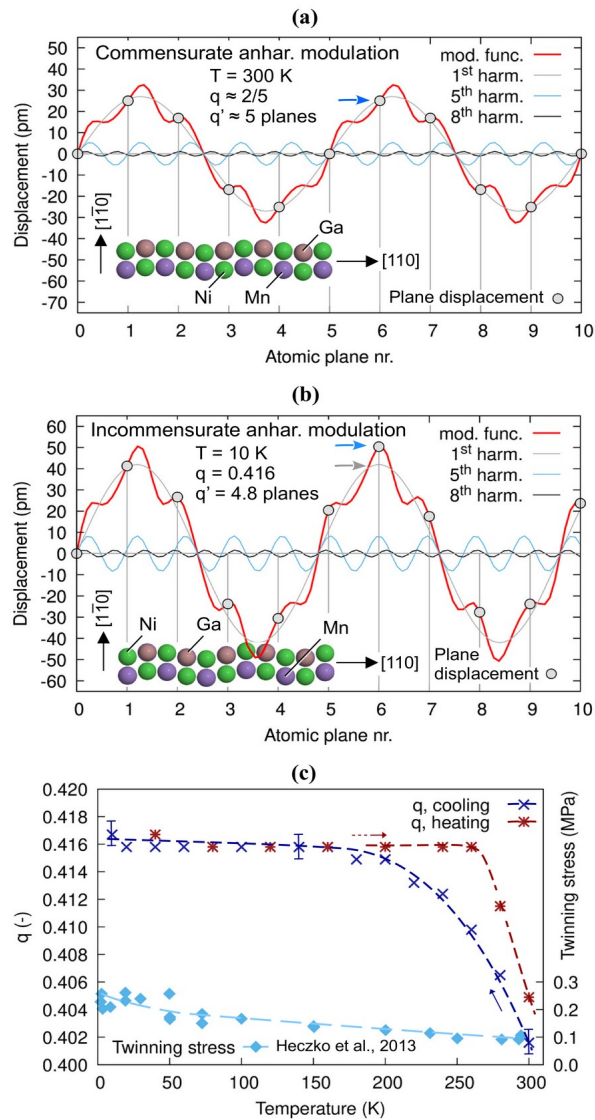